# Hybrid symmetry epitaxy of superconducting Fe(Te,Se) film on a topological insulator

*Xiong Yao, [*,†] Matthew Brahlek,[⊥] Hee Taek Yi,[†] Deepti Jain,[§] Alessandro R. Mazza,[⊥] Myung-Geun Han,[‡] and Seongshik Oh[*,†]*

[†]Center for Quantum Materials Synthesis and Department of Physics & Astronomy, Rutgers, The State University of New Jersey, Piscataway, New Jersey 08854, United States

[⊥] Materials Science and Technology Division, Oak Ridge National Laboratory, Oak Ridge, Tennessee 37831, United States

[§]Department of Physics & Astronomy, Rutgers, The State University of New Jersey, Piscataway, New Jersey 08854, United States

[‡]Condensed Matter Physics and Materials Science, Brookhaven National Laboratory, Upton, New York 11973, United States

*Email: xiong.yao@rutgers.edu

*Email: ohsean@physics.rutgers.edu

Phone: +1 (848) 445-8754 (S.O.)

ABSTRACT

It is challenging to grow an epitaxial four-fold compound superconductor (SC) on six-fold topological insulator (TI) platform due to stringent lattice-matching requirement. Here, we demonstrate that Fe(Te,Se) can grow epitaxially on a TI ($Bi_2Te_3$) layer due to accidental, uniaxial lattice match, which is dubbed as "hybrid symmetry epitaxy". This new growth mode is critical to stabilizing robust superconductivity with $T_C$ as high as 13 K. Furthermore, the superconductivity in this $FeTe_{1-x}Se_x/Bi_2Te_3$ system survives in Te-rich phase with Se content as low as x = 0.03 but vanishes at Se content above x = 0.56, exhibiting a phase diagram that is quite different from that of the conventional Fe(Te,Se) systems. This unique heterostructure platform that can be formed in both TI-on-SC and SC-on-TI sequences opens a route to unprecedented topological heterostructures.

Superconductor (SC)/topological insulator (TI) heterostructure has been intensively investigated since it is considered one of the most promising approaches to implement topological superconductor (TSC) via proximity effect.[1-10] In particular, the proximity-induced TSC provides a number of advantages over the chemical doping method[11-13] in terms of controllability, both theoretically and experimentally. The simplest approach to construct SC/TI heterostructure would be to combine elemental superconductors such as Al, Nb and Pb with the TI layer.[5, 14-18] However, considering that elemental superconductors can easily react with the TI layer and form an interfacial dead layer,[19] a compound superconductor sharing the same anion with the TI layer would be a much better candidate for the SC/TI heterostructure.

Among such compound superconductors, Fe(Te,Se) (or FTS) system stands out in that it exhibits the highest superconducting $T_C$ (well above 10 K) among chalcogenides. However, unlike TI films, which can grow on almost any chemically-stable substrate regardless of lattice matching via van der Waals epitaxy,[20-23] growth of FTS films requires strict lattice matching condition. In particular, because the in-plane symmetry of FTS film is four-fold, the underlying substrate should also have the same four-fold symmetry.[24-26] Accordingly, the only route to superconducting FTS-TI heterostructures over a macroscopic area has been to grow FTS layer first on a lattice-matched substrate and then to grow TI layer on top of the FTS layer using the van der Waals epitaxy mode of the TI layer.[27-33] Considering that certain applications, such as Josephson proximity junctions and SC-TI superlattice,[34] require the SC layer to be on top of the TI layer, it would be highly desirable if a compound superconductor can be grown on top of a TI layer. However, the growth of heterostructure with mixed symmetries are challenging and only be realized in limited cases of elemental metals [35, 36] or in the monolayer-limit.[37-39] So far, no synthesis route to grow an epitaxial and macroscopically contiguous SC layer on a TI has been identified for FTS as well as for any other chalcogenide superconductors.

Here, we provide a solution to this challenging problem by demonstrating that FTS films can grow epitaxially and contiguously on a particular TI, $Bi_2Te_3$, with $T_C$ as high as 13 K, despite the completely different lattice symmetry between the two materials. It turns out that $Bi_2Te_3$ is an unique platform for FTS in that it provides a rare, uniaxial lattice match for FTS, which we call "hybrid symmetry epitaxy" (HSE). The importance of the uniaxial lattice match can be well noted from the contrasting observation that FTS films is found to not grow properly on any other similar materials with different lattice constants such as $Sb_2Te_3$, $Bi_2Se_3$ and $In_2Se_3$. More importantly, unlike the non-superconducting counterpart of FeTe bulk crystal,[40, 41] the superconductivity in $FeTe_{1-x}Se_x/Bi_2Te_3$ preserves in Te-rich phase with Se content as low as x = 0.03, providing a promising platform for investigating the recently discovered Majorana states in FTS [42-45] considering that Te-rich phase is more favorable for topological superconductivity due to its stronger spin-orbit coupling.[43, 46]

We grew the $FTS/Bi_2Te_3$ films by molecular beam epitaxy (MBE): details can be found in the Experimental Section. Considering the different in-plane symmetries of FTS and $Bi_2Te_3$, it is surprising that the reflection high-energy electron diffraction (RHEED) pattern of the FTS overlayer is as good as that of the $Bi_2Te_3$ bottom layer as indicated by bright streaky features in Figure 1a and b, which demonstrates the high-quality epitaxial growth of the FTS layer. The characteristic FTS RHEED pattern appears right after we open the shutters, suggesting a smooth growth transition at the interface, which is also verified by the STEM (scanning tunneling electron microscopy) image in Figure S8. The X-ray diffraction (XRD) 2θ scan shown in Figure 1e also supports the epitaxial growth along c-axis with all the FTS (00m) and $Bi_2Te_3$ (000 3n) peaks clearly observable. The lattice symmetries of these two materials can be verified by the ratio of the spacing of the RHEED streaks in two high-symmetry directions. As indicated by the arrow marks in Figure 1a, the spacing ratio of the wide and narrow streaks is $\sqrt{3}$ for $Bi_2Te_3$ (six-fold), while it is $\sqrt{2}$ for FTS (four-fold) in Figure 1b.

Surprisingly, despite the completely different in-plane lattice constants and symmetries of FTS and $Bi_2Te_3$, the RHEED spacings match almost perfectly in one of the two high symmetry directions as shown in Figure 1a and b. Figure 1c displays the line-cut intensities of these two patterns, confirming almost identical spacings between the two (see Note 1 in Supporting Information Figure S1). This uniaxial lattice matching originates because the in-plane lattice constants of $Bi_2Te_3$ (4.38 Å) and FTS (3.76-3.82 Å, average 3.79 Å) are almost perfectly at the ratio of 2 : $\sqrt{3}$. With this ratio, the two materials with four- and six-fold in-plane symmetries make perfect lattice match along one axis (b-axis) as shown in Figure 1d.

There are two additional notable features in the rotational RHEED patterns of FTS/$Bi_2Te_3$. First, the matching patterns from FTS and $Bi_2Te_3$ are strictly aligned at the same angle. Second, the RHEED pattern of FTS grown on $Bi_2Te_3$ shows a twelve-fold in-plane rotational symmetry (repeating every 30°), even if FTS has four-fold symmetry (repeating every 90°) and Bi$_2$Te$_3$ has six-fold symmetry (repeating every 60°). Figure 1f shows azimuthal XRD φ scans of FTS/Bi$_2$Te$_3$, which is consistent with the twelve-fold RHEED patterns: the twelve-fold small peaks are from twinned FTS (103) planes, while the three large peaks are from the twin-free Bi$_2$Te$_3$ ($10\bar{1}10$) planes: note that even if Bi$_2$Te$_3$ exhibits six-fold symmetry on the surface, the bulk symmetry taking into account the layering sequence is three-fold. The perfect coincidence of Bi$_2$Te$_3$ ($10\bar{1}10$) and FTS (103) planes shown on the right panel of Figure 1f confirms that these two layers are aligned as described in Figure 1d. This can be better understood in the schematic diagram of the twinned four-fold FTS domains aligned with the underlying six-fold Bi$_2$Te$_3$ layer in Figure 1g. The growth of the four-fold FTS on Bi$_2$Te$_3$ follows the uniaxial lattice-matched directions with three crystallographically equivalent orientations labeled by different colors (green, red and blue). The in-plane four-fold symmetry of FTS plus these three orientations generated by Bi$_2$Te$_3$ layer leads to the twelve-fold rotational symmetry in RHEED patterns and XRD φ scans. Figure 1g also illustrates the correspondence between the in-plane angle (φ) of X-ray and the individual XRD peaks in Figure 1f in terms of the three different colors.

Considering that the interaction between FTS and substrates is strong enough to induce strained films,[24, 26] the van der Waals epitaxy is unlikely for the growth mode of FTS. In order to test the role of uniaxial lattice match between FTS and Bi$_2$Te$_3$ for the observed HSE growth, we grew three control samples by employing Sb$_2$Te$_3$, Bi$_2$Se$_3$ and In$_2$Se$_3$ as the bottom layers. These three materials share the same structure and van der Waals nature with $Bi_2Te_3$ but have different in-plane lattice constants.

Figure 2a compares the line-cut intensity of RHEED patterns taken for $Sb_2Te_3$ and FTS/$Bi_2Te_3$. Unlike $Bi_2Te_3$, $Sb_2Te_3$ exhibits a 5.4% uniaxial lattice mismatch with FTS based on the RHEED spacing difference. Figure 2b schematically overlays the FTS lattice on top of $Sb_2Te_3$, representing the 5.4% uniaxial lattice mismatch. In Figure 2c, RHEED patterns of a 20 u.c. FTS film grown on $Sb_2Te_3$ exhibit blurry and spotty streaks with a relative spacing ratio of 1.36 rather than $\sqrt{2}$, and give an in-plane lattice parameter smaller than expected from FTS. The RHEED patterns along with the insulating behavior shown in Figure S2a strongly suggest that the FTS film formed on $Sb_2Te_3$ is not the pure four-fold FTS phase with superconductivity. Moreover, the RHEED pattern characteristic of $Sb_2Te_3$ is still visible after deposition of 20 u.c. FTS, which means that the overlayer does not completely cover the $Sb_2Te_3$ bottom layer, instead clustering into the spotty RHEED feature. This is in stark contrast with the van der Waals epitaxy growth of $Sb_2Te_3$ on top of FeTe.[32]

Figure 2d compares the RHEED spacings of $Bi_2Se_3$ and FTS/$Bi_2Te_3$, which show uniaxial lattice mismatch of 7.7%, as schematically laid out in Figure 2e. With this large lattice mismatch, FTS does not grow in its four-fold symmetry on $Bi_2Se_3$ and instead grows with six-fold in-plane symmetry as shown in the RHEED pattern of Figure 2f: Figure S2b also exhibits degraded superconducting properties with quite broad transition. We also grew a Te-free FeSe film on $Bi_2Se_3$, and it similarly exhibits a six-fold in-plane symmetry with slightly smaller in-plane lattice parameter than $Bi_2Se_3$ as shown in Figure S3b. Furthermore, the RHEED pattern of $Bi_2Se_3$ is still visible after 20 u.c. deposition of FeSe on top, implying that the $Bi_2Se_3$ bottom layer is not even fully covered by the FeSe film grown on top.

Lastly, we tried FTS growth on $In_2Se_3$. The uniaxial lattice mismatch between FTS and $In_2Se_3$ is 9.5% based on the RHEED spacings in Figure 2g and the corresponding lattice layout is shown in Figure 2h. As shown in Figure 2i, the RHEED pattern of FTS grown on $In_2Se_3$ exhibits streaks with four-fold in-plane symmetry. However, multiple online/offline spots appear in the RHEED pattern, implying that FTS grows in 3D islands on top of $In_2Se_3$. Further, the transport property shown in Figure S2c also exhibits poor superconductivity in which the zero-resistance state is absent.

In contrast to these control samples, FTS films grown on $Bi_2Te_3$ not only exhibit the structurally-clean four-fold phase, but also exhibit clear superconducting properties as shown in Figure 3 and 4. $T_C^{onset}$ of a $FeTe_{0.88}Se_{0.12}/Bi_2Te_3$ film as determined by the temperature-dependent resistance in Figure 3a is 12.7 K. As shown in Figure 3b, the Hall resistance measured at 20 K after subtraction of the $Bi_2Te_3$ contribution (for details see Supporting Information S5) gives an n-type carrier density of $1.54 \times 10^{20}/cm^3$, which is two orders smaller than those in previously-reported FTS films but comparable to those in superconducting FeSe bulk crystals.[47, 48] The vastly differing carrier densities among superconducting FTS samples is likely due to the presence of both electron and hole pockets in the FTS system.[49] Understanding how each of these pockets interact with the TI layer in a different way will be critical to fully utilize these SC/TI heterostructures. Figure 3c shows samples that have fixed thickness of $Bi_2Te_3$ but different thickness of FTS. With decreasing thickness, the resistance shows a small shift on $T_C^{onset}$ but an obvious shift on $T_C^0$, leading to a broadened superconducting transition.

Figure 3d,e give the temperature-dependent resistance of a $FeTe_{0.97}Se_{0.03}$(35 u.c.)/$Bi_2Te_3$ sample under magnetic fields, perpendicular and parallel to the ab plane, respectively. The upper critical fields of $FeTe_{0.97}Se_{0.03}$ with varying thickness grown on $Bi_2Te_3$, $H_{C2}^{\perp}(T)$ and $H_{C2}^{//}(T)$, defined as the point where resistance reaches 50% of the normal resistance, are presented in Fig. 3f. If orbital component of the superconducting condensate dominates the magnetic-response of the superconducting films, the Ginzburg–Landau (GL) theory dictates that $H_{C2}^{\perp}(T) = \frac{\Phi_0}{2\pi\xi_{GL}(0)^2}(1 - \frac{T}{T_C})$ and $H_{C2}^{//}(T) = \frac{\sqrt{12}\Phi_0}{2\pi\xi_{GL}(0)d}(1 - \frac{T}{T_C})^{\frac{1}{2}}$, where $\Phi_0$ is the flux quantum, d is the superconducting thickness and $\xi_{GL}(0)$ is the zero-temperature GL coherence length. One of the notable features of these GL equations is the linear temperature dependence of $H_{C2}^{\perp}(T)$, which is universally observed on many superconductors near $T_C$, where the GL theory becomes most reliable. In particular, this linear $H_{C2}^{\perp}(T)$ dependence showed up prominently on the $Bi_2Te_3$/FeTe heterostructures,[27, 32] where the superconductivity is believed to reside only at the interfaces. In our films as shown in Figure 3f, $H_{C2}^{\perp}(T)$ is linear in the thin limit (5 u.c.), but becomes sublinear as the film gets thicker. Furthermore, the anisotropy of the critical fields, defined as $H_{C2}^{//}(T)/H_{C2}^{\perp}(T)$, also becomes weaker as the FTS film gets thicker, getting comparable to that of the bulk, according to Figure 3d, e and S9. These findings suggest that even though we cannot rule out the presence of interfacial superconductivity, our FTS/$Bi_2Te_3$ platform clearly has strong bulk

contribution: whether the bulk and interfacial superconductivities coexist or not remains an open question and needs more in-depth studies.

Figure 4a,b show the transport properties of the FeTe$_{1-x}$Se$_x$/Bi$_2$Te$_3$ samples with different Se content, which is estimated by RBS (Rutherford backscattering spectroscopy) as described in Note 2 in Supporting Information Figure S1. With increasing Se content, $T_C$ reaches a maximum at x = 0.28, and then gradually decreases and vanishes for x = 0.56 and 1. We summarized $T_C^{onset}$ values of all the samples in Figure 4c, together with $T_C^{onset}$ of FTS bulk crystals and FTS/CaF$_2$ films.[26,40,41] It is notable that superconductivity is preserved up to the maximum Te-rich phase (x = 0.03), which is known to be non-superconducting in bulk crystals,[40,41] but absent on the Se-rich side. In particular, the very absence of superconductivity in FeSe, which is a well-established superconductor in both films and bulk crystals,[25,48] is surprising. One possibility is the charge transfer effect, which was believed to induce the interfacial superconductivity in Bi$_2$Te$_3$/FeTe.[30,33] The difference in band alignment of FeSe vs FeTe relative to Bi$_2$Te$_3$ could lead to widely varying charge transfer to and from the TI layer (See Supporting Information Figure S5 and S6), possibly shifting the superconducting dome on the phase diagram until completely suppressing superconductivity in FeSe while enhancing it in FeTe. The complete removal of excess Fe in our samples, as can be seen from the RBS results in Table S1, could be another possible reason for the emergence of superconductivity in FeTe since it has been reported that excess Fe is deleterious for the superconductivity in both FTS bulk crystals and TI/FeTe films.[32,50] A structural effect at the interface could also help develop the superconductivity, as proposed in a recent scanning tunneling microscopy (STM) study of a monolayer FeTe film grown on $Bi_2Te_3$.[39] However, neither of these two mechanisms can explain the absence of superconductivity in Se-rich samples. Further studies are required to fully resolve this mystery.

In conclusion, we discovered a new epitaxy mode, HSE, in a unique heterostructure platform of FTS/$Bi_2Te_3$. Despite the completely different lattice symmetry between the two materials, the uniaxial lattice match allows growth of epitaxial and superconducting FTS films on $Bi_2Te_3$. Comparison with other control samples grown on structurally similar bottom layers confirms the essential role of the uniaxial lattice match for HSE. This discovery opens new routes to utilizing HSE to combine materials previously overlooked under the conventional paradigm of epitaxy. In particular, the superconductivity in FTS/$Bi_2Te_3$ displays a completely different phase diagram than those of conventional FTS systems, suggesting non-trivial role of the underlying $Bi_2Te_3$ layer. Moreover, as demonstrated in Supporting Information Figure S7, FTS-$Bi_2Te_3$ is a unique TI-SC heterostructure that can be formed in TI-on-SC as well as SC-on-TI sequences, so this platform may also provide a route to the elusive Weyl superconductors in the form of SC/TI superlattices as theoretically proposed several years ago.[34]

**Growth method**: All samples were grown on 10 mm × 10 mm $Al_2O_3$ (0001) substrates by a custom built SVTA MOS-V-2 MBE system with a base pressure of low $10^{-10}$ Torr. 99.999% pure elemental Bi, Sb, Te, Se, In and Fe sources were thermally evaporated using Knudsen cells for film growth. All the source fluxes were calibrated in situ by quartz crystal micro-balance and ex situ by RBS. Substrates were cleaned ex situ by 5 minutes exposure to UV-generated ozone and in situ by heating to 750 ºC under oxygen pressure of $1 \times 10^{-6}$ Torr for ten minutes. For $Bi_2Te_3$ ($Sb_2Te_3$/$Bi_2Se_3$/$In_2Se_3$) growth, an initial 3 quintuple layer (QL) working as seed layer were deposited at a relatively low temperature of 240 °C, followed by the growth of 7 QL at 300 °C. Then the shutters of Fe, Te and Se sources with pre-adjusted fluxes were opened at the same growth temperature of 300 °C for FTS deposition.

**Transport measurement:** All transport measurements were performed by manually pressing four indium wires on the sample. The transport measurements were performed in both a closed-cycle cryostat (6 K) and a Quantum Design Physical Property Measurement System (PPMS; 2 K). There could be slight calibration difference between the thermometers in two systems. Raw data of $R_{xx}$ and $R_{xy}$ were properly symmetrized and antisymmetrized respectively.

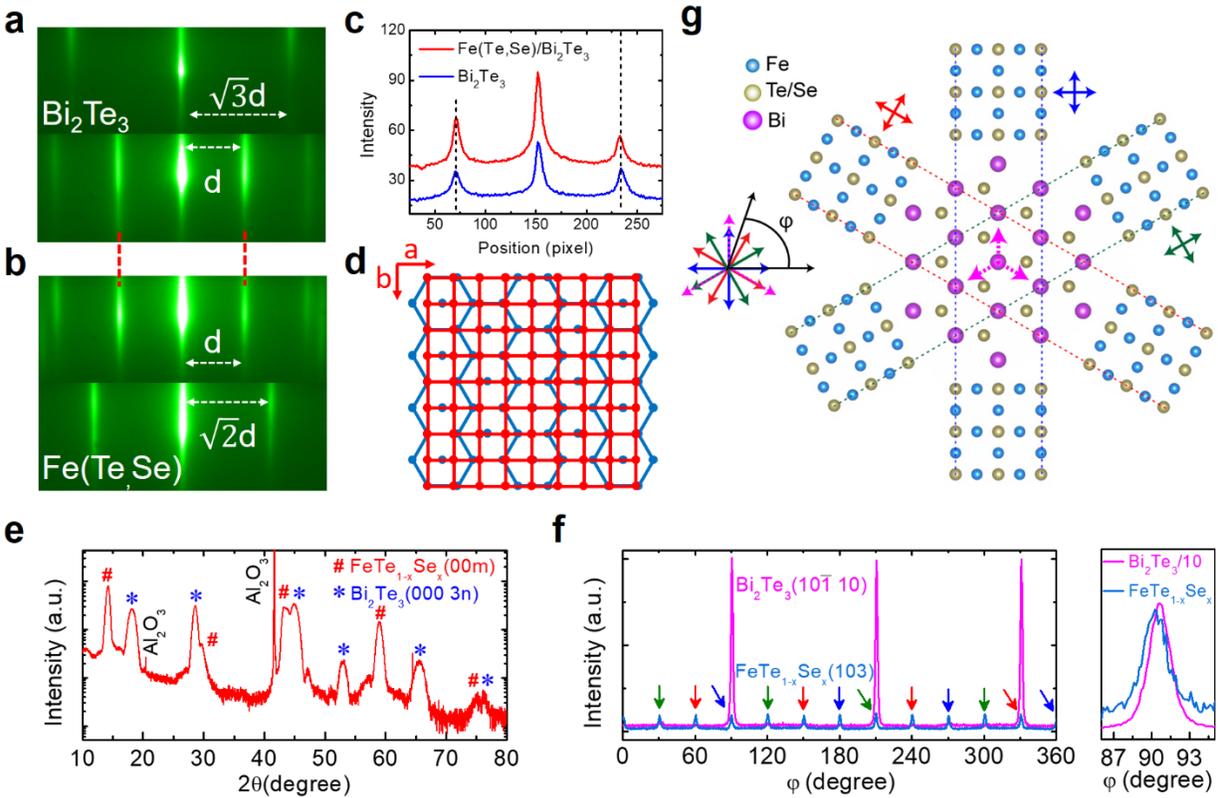

Figure 1: Characterizations of the FTS/Bi$_2$Te$_3$ heterostructure. (a and b) Reflection high-energy electron diffraction (RHEED) patterns of (a) Bi$_2$Te$_3$ and (b) FTS/Bi$_2$Te$_3$. The arrow marks indicate the RHEED streak spacings. The red dash guidelines indicate that the RHEED spacings are the same. (c) The line-cut intensities of the narrow streaks in (a) and (b). (d) Schematic of FTS (red) lattice overlaid on top of Bi$_2$Te$_3$ (blue) lattice: the dots represent Te(Se) atoms. (e) XRD 2θ scan of a FTS(40 u.c.)/Bi$_2$Te$_3$(10 nm) film. (f) In-plane XRD φ scans of the same sample in (e). The different color labels of FTS (103) peaks indicate twins with three orientations. (g) Schematic diagram of azimuthal XRD φ scans for Bi$_2$Te$_3$ and twinned FTS lattices: the three crystallographically equivalent orientations of FTS are labeled by three different colors. The dash

guidelines indicate the uniaxial lattice match between FTS and $Bi_2Te_3$. The X ray directions shown in φ coordinate system are consistent with the peak labels in (f).

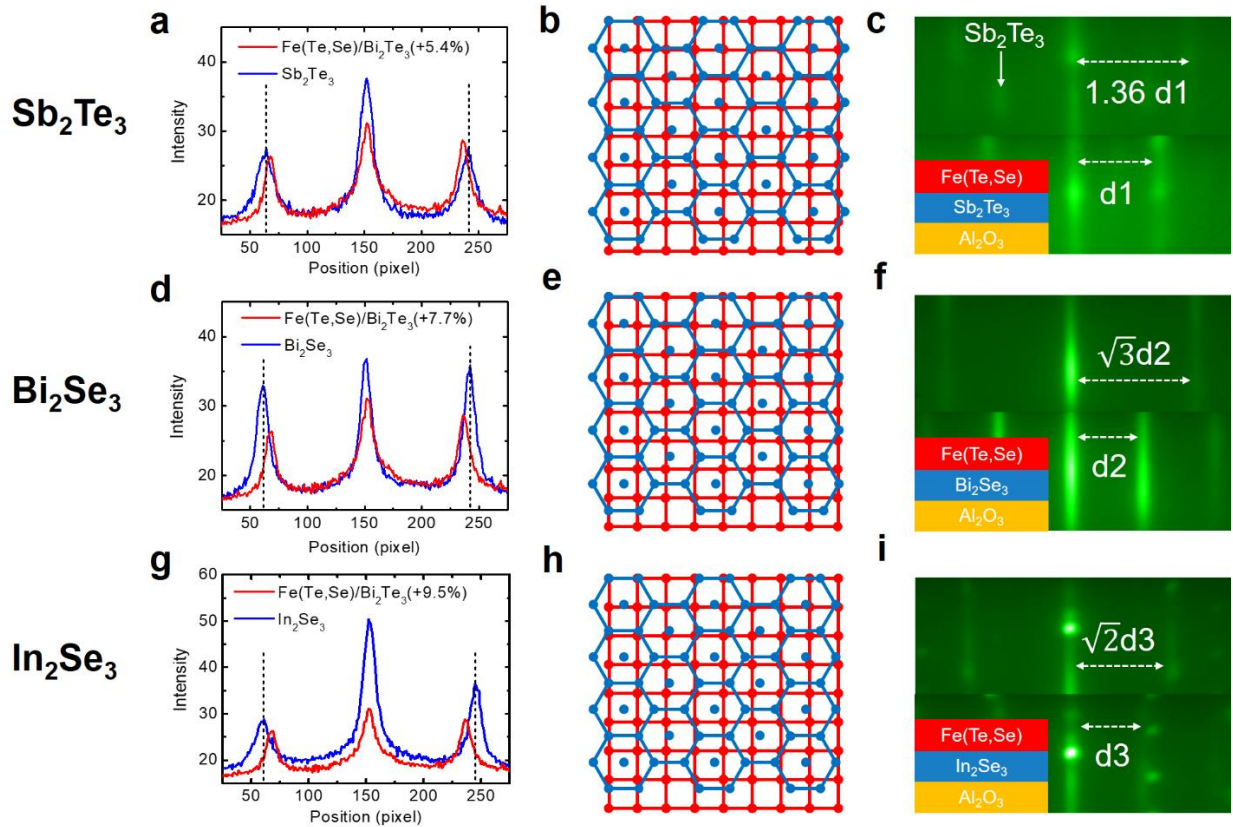

Figure 2: Three control samples for testing the uniaxial lattice match. (a, d and g) Line-cut intensities of the narrow RHEED streaks from FTS/$Bi_2Te_3$ and (a) $Sb_2Te_3$, (d) $Bi_2Se_3$ and (g) $In_2Se_3$. (b, e and h) Schematics of overlaid lattices for (b) FTS/$Sb_2Te_3$, (e) FTS/$Bi_2Se_3$ and (h) FTS/$In_2Se_3$. (c, f and i) RHEED patterns of 20 u.c. FTS films grown on (c) $Sb_2Te_3$, (f) $Bi_2Se_3$ and (i) $In_2Se_3$.

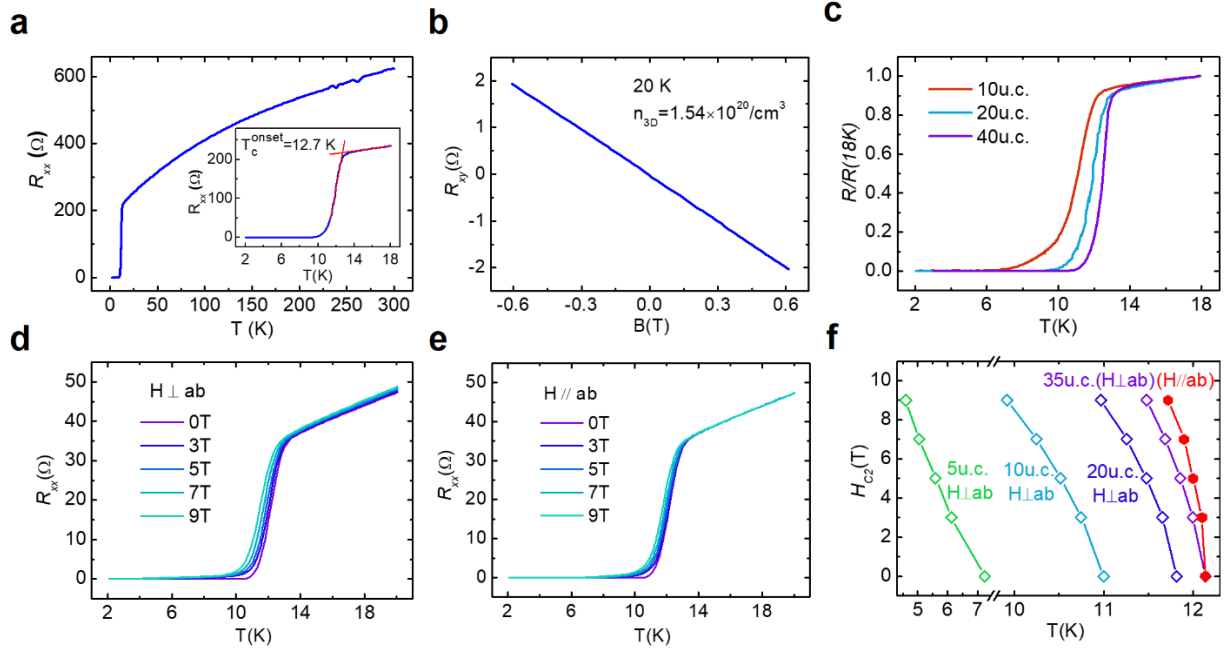

Figure 3: Transport properties for FeTe$_{1-x}$Se$_x$/Bi$_2$Te$_3$ (10 nm) films. (a) Temperature-dependent longitudinal resistance of a 20 u.c. film with x=0.12 from 300 K to 2 K. Inset shows the enlarged plot from 2 K to 18 K. Intersection of linear extrapolations from normal-state and superconducting transition regions gives $T_C^{onset}$ of 12.7 K. (b) Field-dependent Hall resistance of the same sample in (a) after subtracting the Bi$_2$Te$_3$ contribution: details are included in Supporting Information Figure S5. The 3D carrier density shown in the inset is obtained with the unit cell thickness of FTS determined as 6.19 Å from the XRD 2θ scan. (c) Normalized longitudinal resistance of 10 u.c, 20 u.c. and 40 u.c. FTS films with x = 0.12 grown on 10 nm Bi$_2$Te$_3$. (d, e) Temperature-dependent longitudinal resistance of a 35 u.c. film with x = 0.03 under varying magnetic fields (d) perpendicular and (e) parallel to the ab plane. (f) $H_{C2}^{\perp}(T)$ of 5 u.c., 10 u.c., 20 u.c. and 35 u.c.

FeTe$_{0.97}$Se$_{0.03}$ grown on Bi$_2$Te$_3$ and $H_{C2}^{//}(T)$ of 35u.c. FeTe$_{0.97}$Se$_{0.03}$/Bi$_2$Te$_3$, determined by 50% $R_n$.

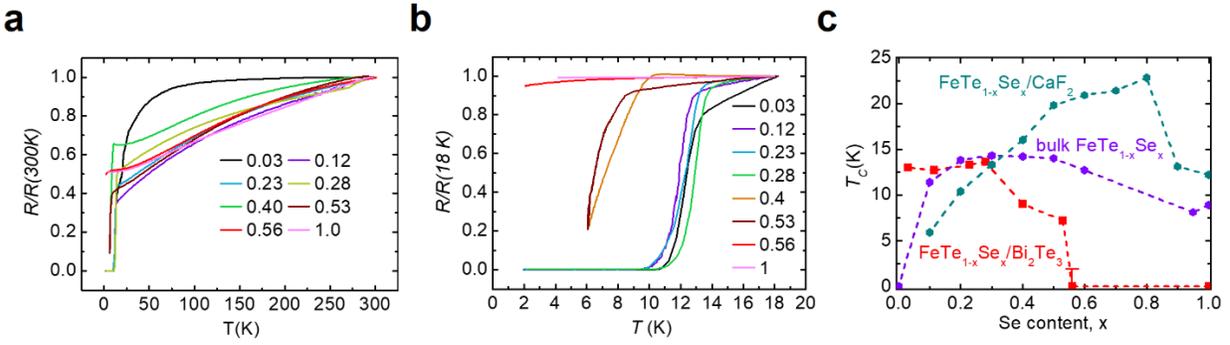

Figure 4: Superconducting properties of FeTe$_{1-x}$Se$_x$/Bi$_2$Te$_3$(10 nm) films as a function of Se content. (a and b) Normalized longitudinal resistance of FeTe$_{1-x}$Se$_x$/Bi$_2$Te$_3$ films with varying Se content. (c) Comparison of $T_C^{onset}$ with FeTe$_{1-x}$Se$_x$/CaF$_2$ films and FeTe$_{1-x}$Se$_x$ bulk crystals. Here the thickness of FeTe$_{1-x}$Se$_x$ layers is 20 u.c. except x = 0.03, which is 35 u.c.

ASSOCIATED CONTENT

**Supporting Information**.

Summary of Rutherford backscattering spectroscopy (RBS) simulation results for the FTS/$Bi_2Te_3$ samples, reflection high-energy electron diffraction (RHEED) and RBS data for the FTS/$Bi_2Te_3$ heterostructures, transport results of the control samples, RHEED patterns of the FeSe/$Bi_2Se_3$ control samples, thickness-dependent transport results of FTS/$Bi_2Te_3$, Hall resistance of FTS/$Bi_2Te_3$, Hall resistance of FeSe/$Bi_2Te_3$, characterizations of a $Bi_2Te_3$/FeSe/$Bi_2Te_3$ sample, cross-sectional characterizations of a FeTe$_{0.97}$Se$_{0.03}$/$Bi_2Te_3$ sample, superconducting properties of FeTe$_{0.97}$Se$_{0.03}$/$Bi_2Te_3$ sample with varying thickness. This material is available free of charge via the Internet at http://pubs.acs.org.

AUTHOR INFORMATION

Corresponding Authors

*E-mail: xiong.yao@rutgers.edu

*E-mail: ohsean@physics.rutgers.edu

Author Contributions

X.Y. and S.O. conceived the experiments. X.Y., H.Y. and D.J. grew the thin films. X.Y. performed the transport measurements and analyzed the data with S.O. M.B. and A.M. performed the XRD measurements. M.H. performed the STEM measurements. X.Y. and S.O. wrote the manuscript with contributions from all authors.

Notes

The authors declare no competing financial interest.

ACKNOWLEDGMENT

This work is supported by National Science Foundation's DMR2004125, Army Research Office (ARO) Grant No. W911NF-20-1-0108, the center for Quantum Materials Synthesis (cQMS), funded by the Gordon and Betty Moore Foundation's EPiQS initiative through grant GBMF6402, and the U.S. Department of Energy, Office of Science, National Quantum Information Science Research Centers and the Basic Energy Sciences, Materials Sciences and Engineering Division. The work at Brookhaven National Laboratory is supported by the Materials Science and Engineering Divisions, Office of Basic Energy Sciences of the U.S. Department of Energy under contract no. DESC0012704. FIB use at the Center for Functional Nanomaterials, Brookhaven National Laboratory is acknowledged.

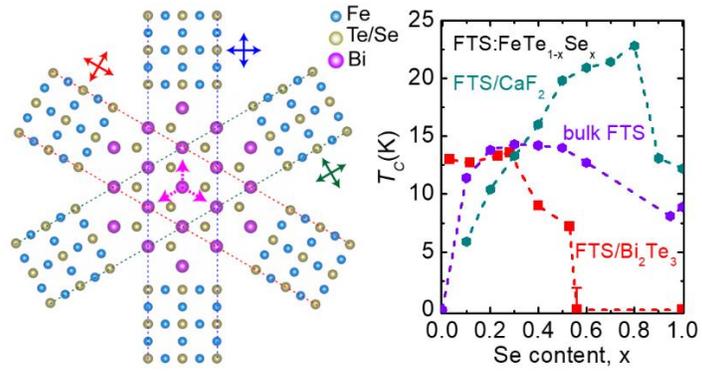

For TOC only

*Supporting Information:*

# Hybrid symmetry epitaxy of superconducting Fe(Te,Se) film on a topological insulator

*Xiong Yao, \*,† Matthew Brahlek,⊥ Hee Taek Yi,† Deepti Jain,§ Alessandro R. Mazza,⊥ Myung-Geun Han,‡ and Seongshik Oh\*,†*

†Center for Quantum Materials Synthesis and Department of Physics & Astronomy, Rutgers, The State University of New Jersey, Piscataway, New Jersey 08854, United States

⊥ Materials Science and Technology Division, Oak Ridge National Laboratory, Oak Ridge, Tennessee 37831, United States

§Department of Physics & Astronomy, Rutgers, The State University of New Jersey, Piscataway, New Jersey 08854, United States

‡Condensed Matter Physics and Materials Science, Brookhaven National Laboratory, Upton, New York 11973, United States

*Email: xiong.yao@rutgers.edu

*Email: ohsean@physics.rutgers.edu

Phone: +1 (848) 445-8754 (S.O.)

## 1. Summary of Rutherford backscattering spectroscopy (RBS) simulation results for the Fe(Te,Se)/Bi$_2$Te$_3$ samples

| Sample | Bi$_2$Te$_3$ | | Fe(Te,Se) | | | Fe : Te : Se | x |
|---|---|---|---|---|---|---|---|
| | Bi | Te | Fe | Te | Se | | |
| A | 0.98 | 1.47 | 5.01 | 4.87 | 0.17 | 1 : 0.97 : 0.03 | 0.03 |
| B | 1.25 | 1.88 | 2.64 | 2.43 | 0.32 | 1 : 0.92 : 0.12 | 0.12 |
| C | 1.20 | 1.79 | 2.61 | 2.01 | 0.59 | 1 : 0.77 : 0.23 | 0.23 |
| D | 1.24 | 1.86 | 2.53 | 1.80 | 0.71 | 1 : 0.71 : 0.28 | 0.28 |
| E | 1.19 | 1.78 | 2.72 | 1.66 | 1.08 | 1 : 0.61 : 0.40 | 0.40 |
| F | 1.21 | 1.81 | 2.62 | 1.26 | 1.39 | 1 : 0.48 : 0.53 | 0.52 |
| G | 1.22 | 1.83 | 2.46 | 1.08 | 1.38 | 1 : 0.44 : 0.56 | 0.56 |

Table S1: Chemical compositions for all the Fe(Te,Se)/Bi$_2$Te$_3$ samples based on RBS results. The atom numbers are in units of $10^{16}$/cm$^2$, and x is the Se content in the form of Te$_{1-x}$Se$_x$.

## 2. Reflection high-energy electron diffraction (RHEED) and RBS data for the Fe(Te,Se)/Bi$_2$Te$_3$ heterostructures

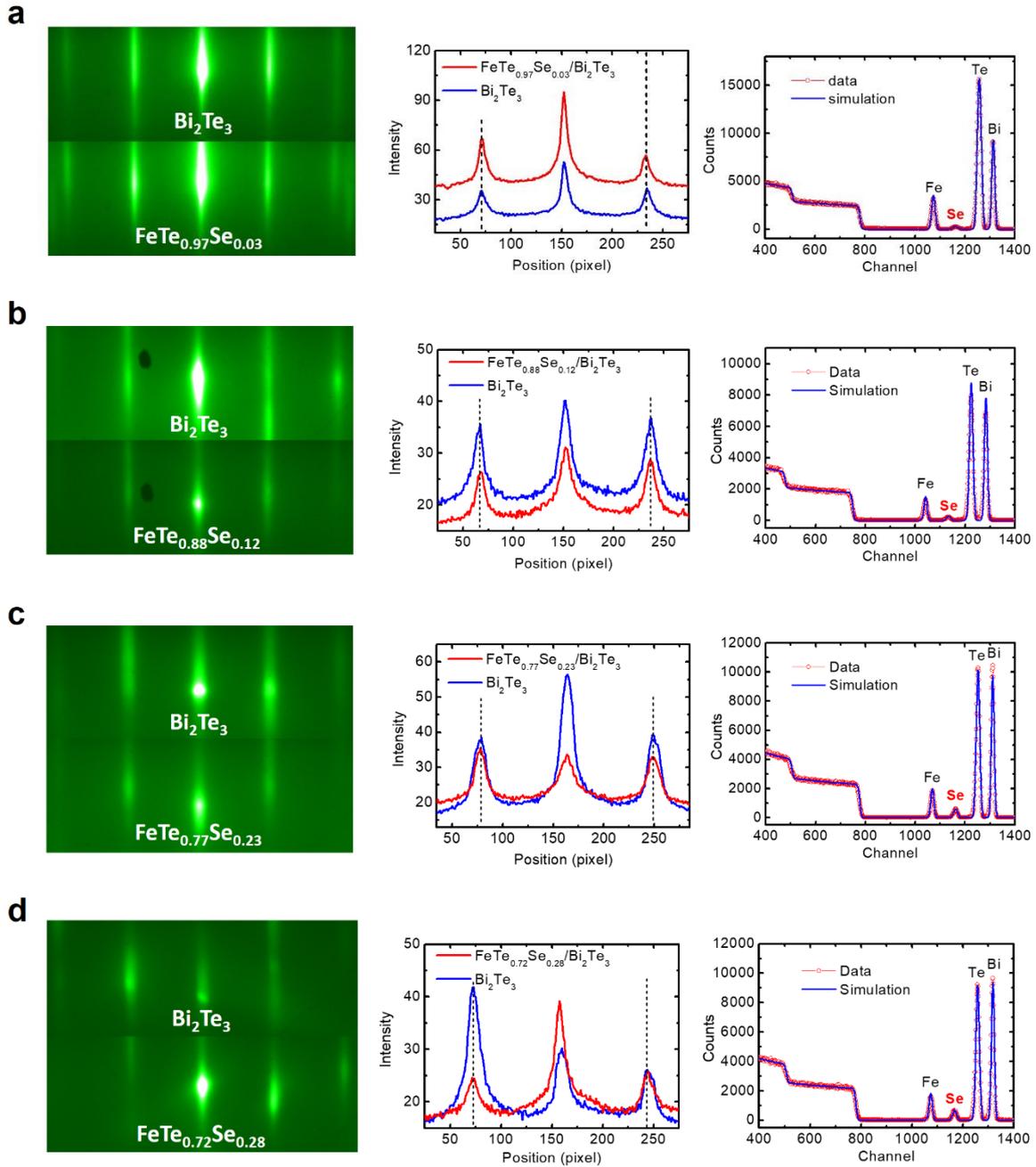

**e**

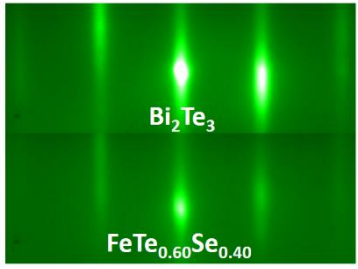 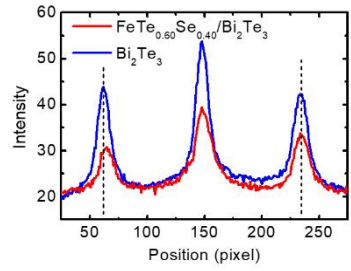 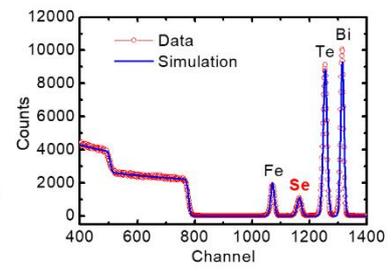

**f**

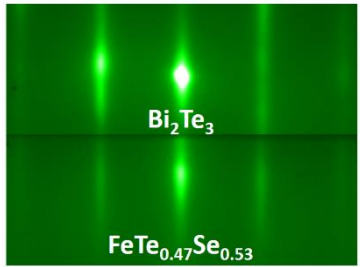 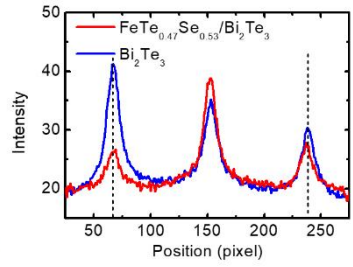 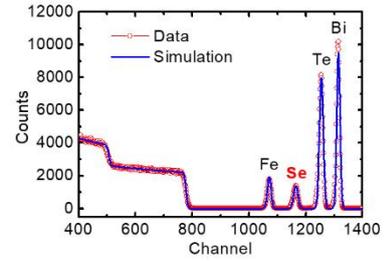

**g**

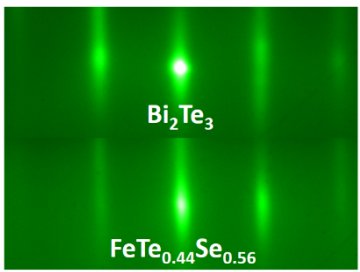 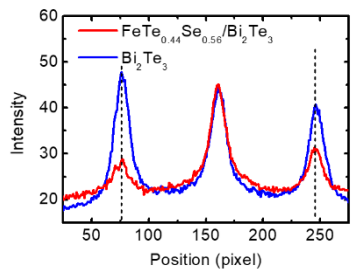 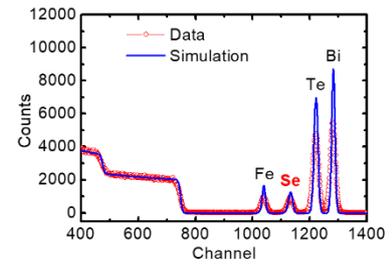

**h**

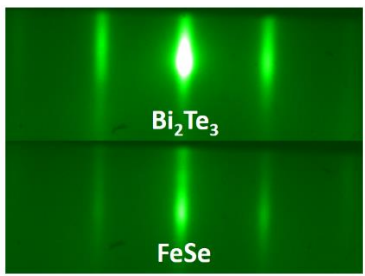 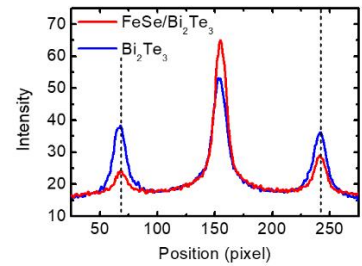

Figure S1: Matching RHEED patterns of $Bi_2Te_3$(10 QL) and Fe(Te,Se)(20 u.c.)/$Bi_2Te_3$, along with line-cut intensities of the RHEED streaks and RBS data with simulations for (a) x = 0.03, (b) x = 0.12, (c) x = 0.23, (d) x = 0.28, (e) x = 0.40, (f) x = 0.53, (g) x = 0.56, (h) x = 1.

Note 1: Lattice constant of Fe(Te,Se) varies from 3.82 Å (FeTe) to 3.76 Å (FeSe), with the average value of 3.79 Å perfectly matching that (4.38 Å) of $Bi_2Te_3$ with the $\sqrt{3}/2$ ratio. In principle, RHEED spacing comparison can be used to keep track of such lattice constant change. However, as shown in the series of RHEED pattern comparisons in Figure S1, within our RHEED resolution, the RHEED spacings of all the Fe(Te,Se) films are identical to those of the underlying $Bi_2Te_3$ patterns regardless of the Se content, and we cannot tell whether this is because Fe(Te,Se) lattices are all clamped to the $Bi_2Te_3$ lattice or because the spacing difference is beyond the resolution of our RHEED setup.

Note 2: According to our previous RBS measurements, the Bi to Te atoms ratios in $Bi_2Te_3$ layer are always very close to the fixed value of 2:3, so we attribute the rest of Te signal to the Fe(Te,Se) layer in the Fe(Te,Se)/$Bi_2Te_3$ bilayer system.

Note 3: We found that the Se content x is always close to the Se flux to Fe flux ratio rather than Se flux to Te flux ratio. Given the fact that Te flux is usually over ten times more than Fe flux, fine-tuning of the low Se flux is crucial for controlling the Se content in Fe(Te,Se) films. This suggests that Fe tends to bond with Se much more easily than with Te.

## 3. Transport results of the control samples

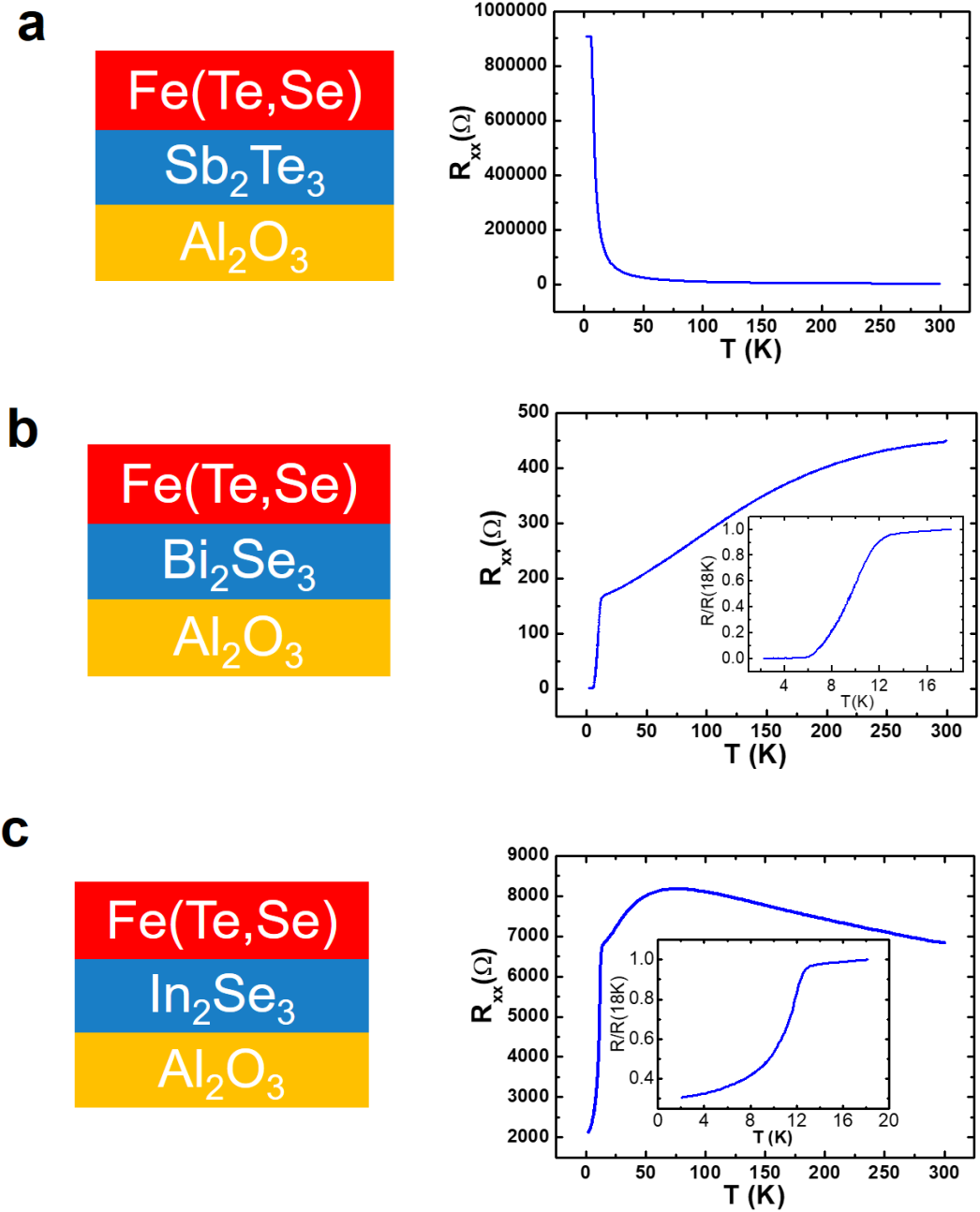

Figure S2: Transport results for (a) Fe(Te,Se)/Sb$_2$Te$_3$, (b) Fe(Te,Se)/Bi$_2$Se$_3$ and (c) Fe(Te,Se)/In$_2$Se$_3$. The thickness of Sb$_2$Te$_3$, Bi$_2$Se$_3$ and In$_2$Se$_3$ are 10 QL, and the thickness of Fe(Te,Se) are 20 u.c..

## 4. RHEED patterns of the FeSe/Bi$_2$Se$_3$ control samples

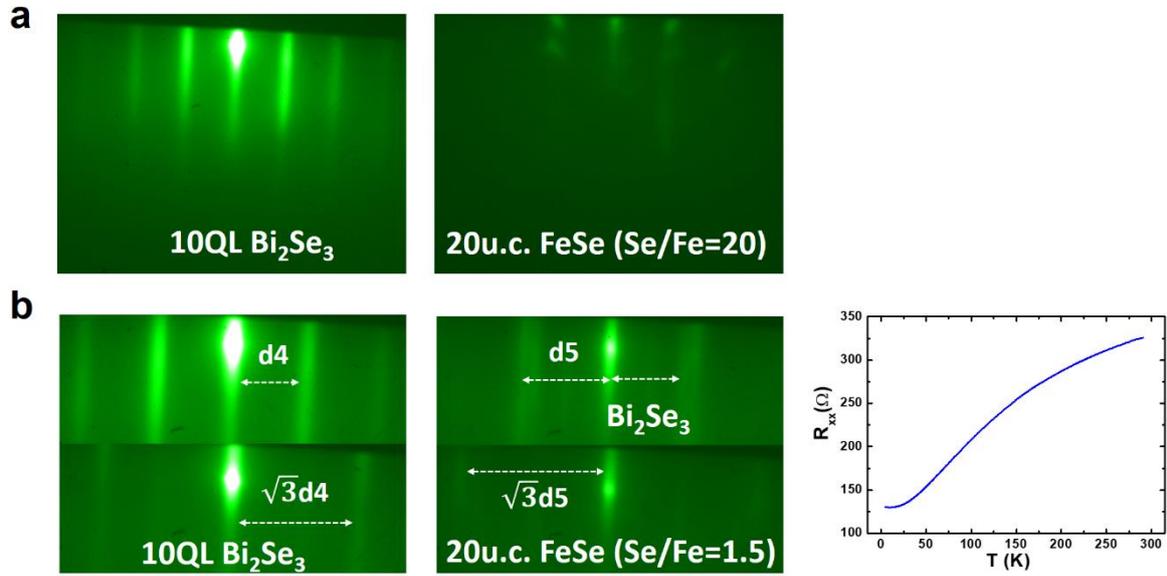

Figure S3: RHEED patterns and transport results of FeSe(20 u.c.)/Bi$_2$Se$_3$(10 QL) films grown in two different ways. (a) Se flux to Fe flux ratio is 20 during the growth of FeSe. RHEED pattern of FeSe shows dark spotty features. (b) Se flux to Fe flux ratio is 1.5 during the growth of FeSe. RHEED patterns of FeSe/Bi$_2$Se$_3$ show coexistence of two sets of patterns. One pattern is consistent with Bi$_2$Se$_3$ RHEED spacing, and the other unknown phase shows a wider spacing with six-fold rotational symmetry. No superconducting transition was observed in any of these films as shown on the right panel.

**5. Thickness-dependent transport results of Fe(Te,Se)/Bi$_2$Te$_3$**

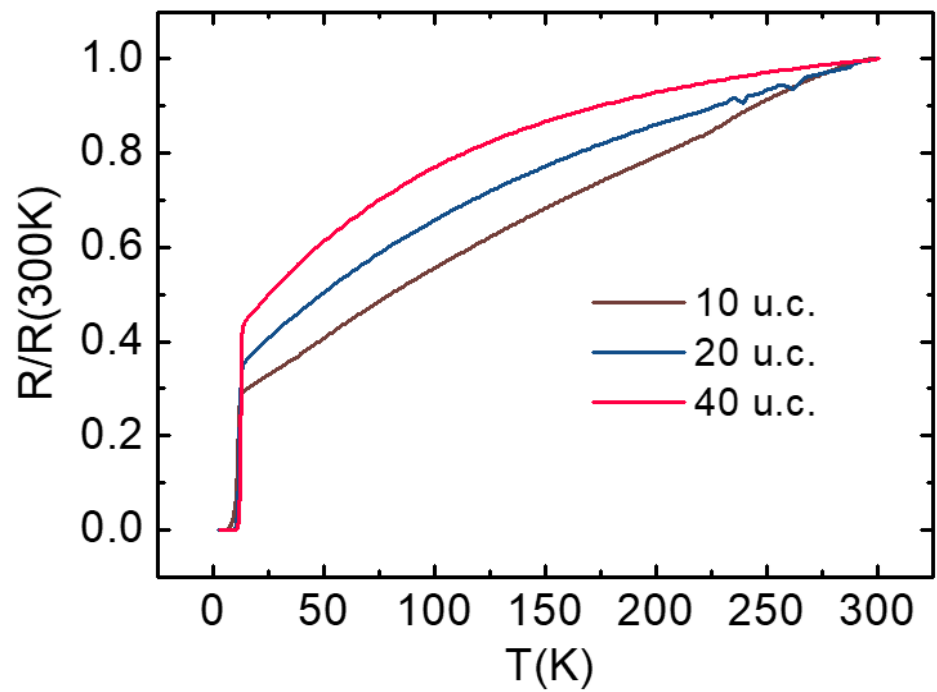

Figure S4: Normalized temperature-dependent longitudinal resistance from 2 K to 300 K for Fe(Te,Se)/Bi$_2$Te$_3$ heterostructures with different thickness of Fe(Te,Se).

## 6. Hall resistance of Fe(Te,Se)/Bi$_2$Te$_3$

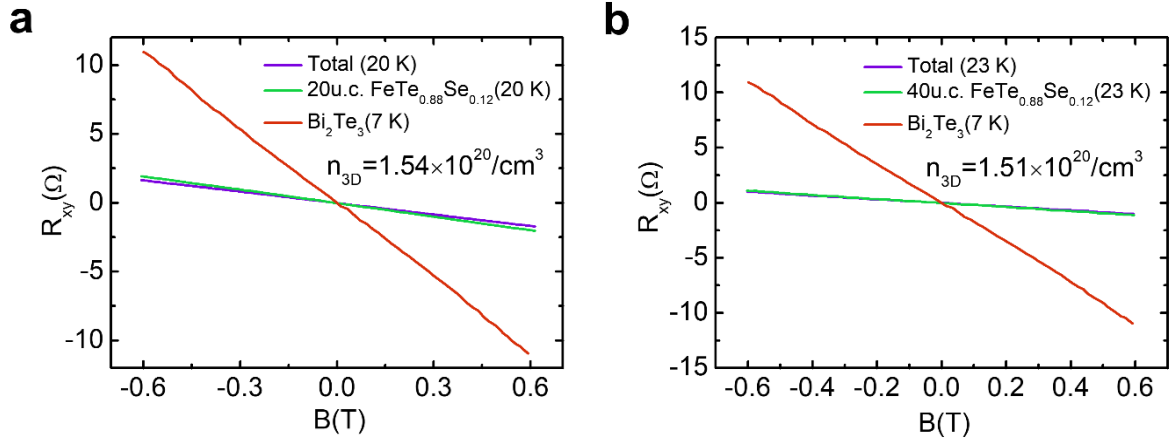

Figure S5: Field-dependent Hall resistance of 20 u.c. and 40 u.c. FeTe$_{0.88}$Se$_{0.12}$ films grown on 10 QL Bi$_2$Te$_3$. The Hall resistance of 10 QL Bi$_2$Te$_3$ was measured at 7 K. The Hall resistance of our Bi$_2$Te$_3$ films are known to be nearly temperature-independent at low temperatures, so we assume that the Hall resistances of Bi$_2$Te$_3$ at 20 K and 23 K are similar to the value at 7 K. Considering that the mobility of the underlying 10 QL Bi$_2$Te$_3$ films is close to that of FeTe$_{0.88}$Se$_{0.12}$/Bi$_2$Te$_3$, we can estimate the Hall resistance of the FeTe$_{0.88}$Se$_{0.12}$ layer using $\frac{1}{R_{xy}(Total)} = \frac{1}{R_{xy}(FeTe_{0.88}Se_{0.12})} + \frac{1}{R_{xy}(Bi_2Te_3)}$. The unit cell thickness value of FeTe$_{0.88}$Se$_{0.12}$ used for carrier density calculation is 6.19 Å from XRD 2θ scan. The 3D carrier density obtained this way is $1.54 \times 10^{20}$/cm$^3$ for 20u.c. FeTe$_{0.88}$Se$_{0.12}$ and $1.51 \times 10^{20}$/cm$^3$ for 40u.c. FeTe$_{0.88}$Se$_{0.12}$, both being n-type. For comparison, 3D carrier density of the 10 QL Bi$_2$Te$_3$ layer is $3.44 \times 10^{19}$/cm$^3$, also n-type.

## 7. Hall resistance of FeSe/Bi$_2$Te$_3$

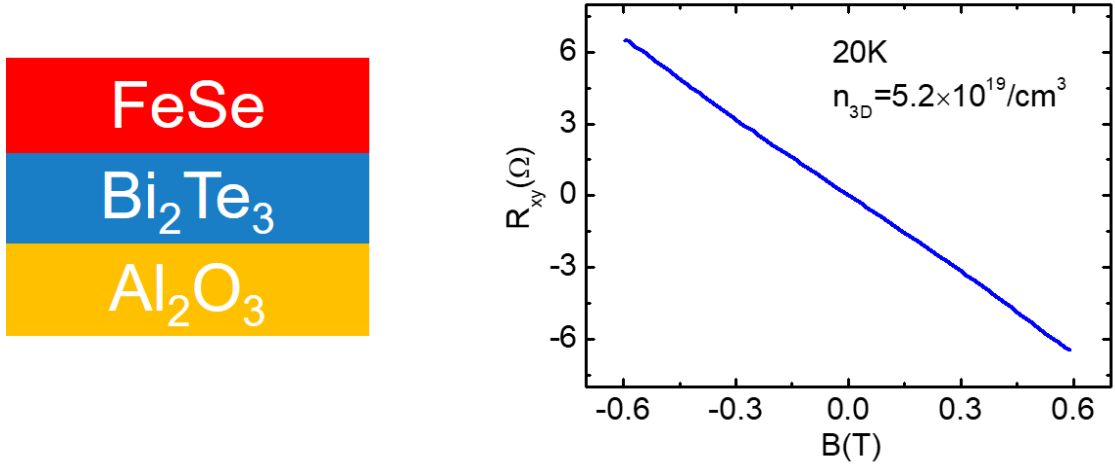

Figure S6: Field-dependent Hall resistance of a 20 u.c. FeSe film grown on 10 QL Bi$_2$Te$_3$. Unlike the FeTe$_{0.88}$Se$_{0.12}$/Bi$_2$Te$_3$ films, the mobility of the FeSe/Bi$_2$Te$_3$ film turns out to be much lower than that of Bi$_2$Te$_3$, so it is hard to extract the Hall resistance of FeSe layer alone. But the total carrier density of FeSe/Bi$_2$Te$_3$, which is $5.2 \times 10^{19}$/cm$^3$, sets the upper bound of the carrier density for the FeSe layer. This value is substantially lower than that of Fe(Te,Se) films on Bi$_2$Te$_3$ as well as that of FeSe bulk crystals, suggesting that the charge transfer effect is likely the main origin for not only the lack of superconductivity in FeSe/Bi$_2$Te$_3$ but also the robust superconductivity in the Te-rich Fe(Te,Se)/Bi$_2$Te$_3$ films.

## 8. Characterizations of a Bi$_2$Te$_3$/FeSe/Bi$_2$Te$_3$ sample

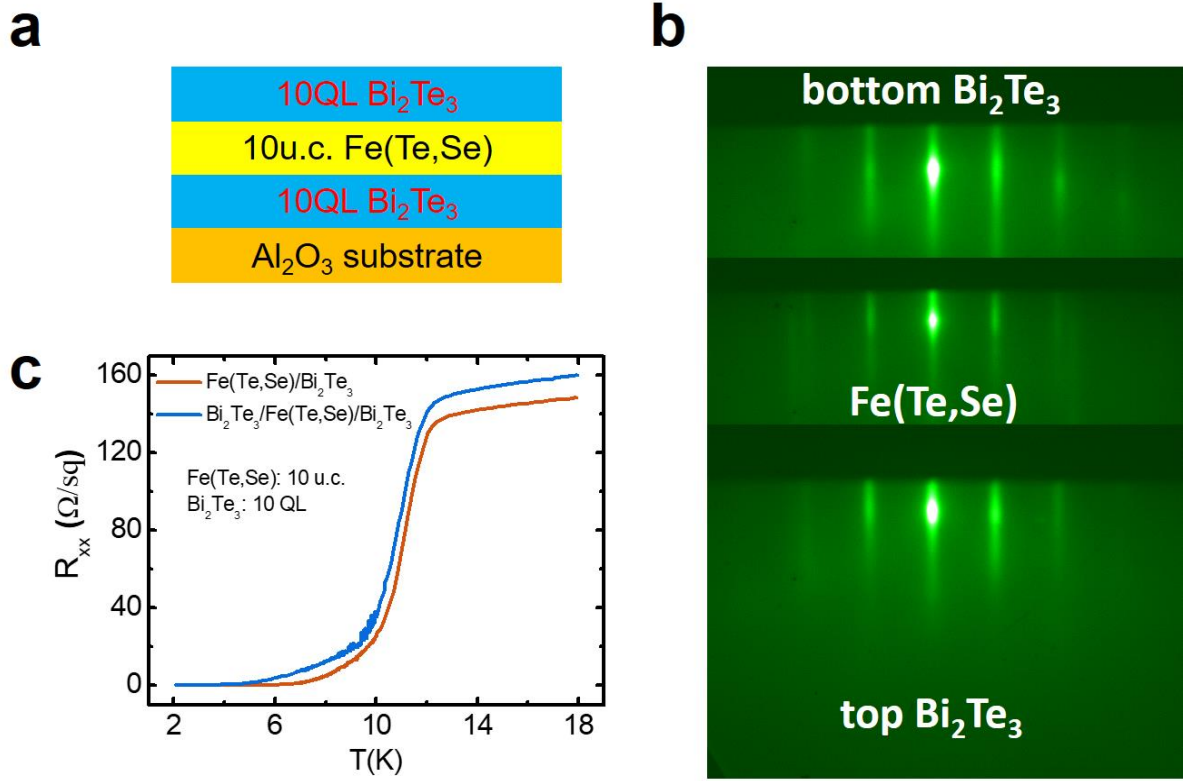

Figure S7: (a) The schematic of the Bi$_2$Te$_3$/Fe(Te,Se)/Bi$_2$Te$_3$ sample. (b) The RHEED patterns of the bottom Bi$_2$Te$_3$ layer, Fe(Te,Se) layer and top Bi$_2$Te$_3$ layer. (c) The comparison of transport properties of Bi$_2$Te$_3$/Fe(Te,Se)/Bi$_2$Te$_3$ and Bi$_2$Te$_3$/Fe(Te,Se)/Bi$_2$Te$_3$. The thickness of Bi$_2$Te$_3$ are all 10 QL, and the thickness of Fe(Te,Se) is 10 u.c.

## 9. Cross-sectional characterizations of a FeTe$_{0.97}$Se$_{0.03}$/Bi$_2$Te$_3$ sample

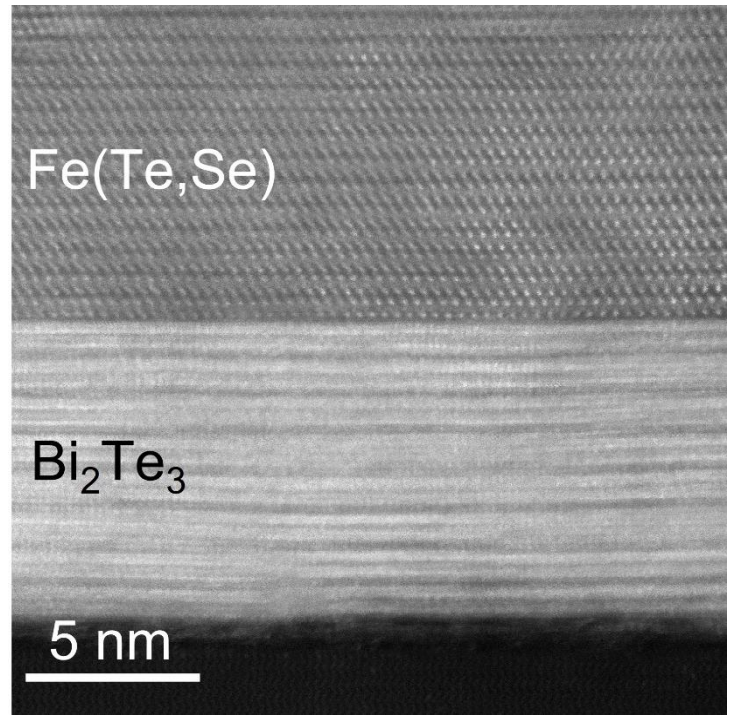

Figure S8: High-angle annular dark-field scanning tunneling electron microscopy (HAADF-STEM) image for a FeTe$_{0.97}$Se$_{0.03}$/Bi$_2$Te$_3$ sample.

## 10. Superconducting properties of FeTe$_{0.97}$Se$_{0.03}$/Bi$_2$Te$_3$ sample with varying thickness

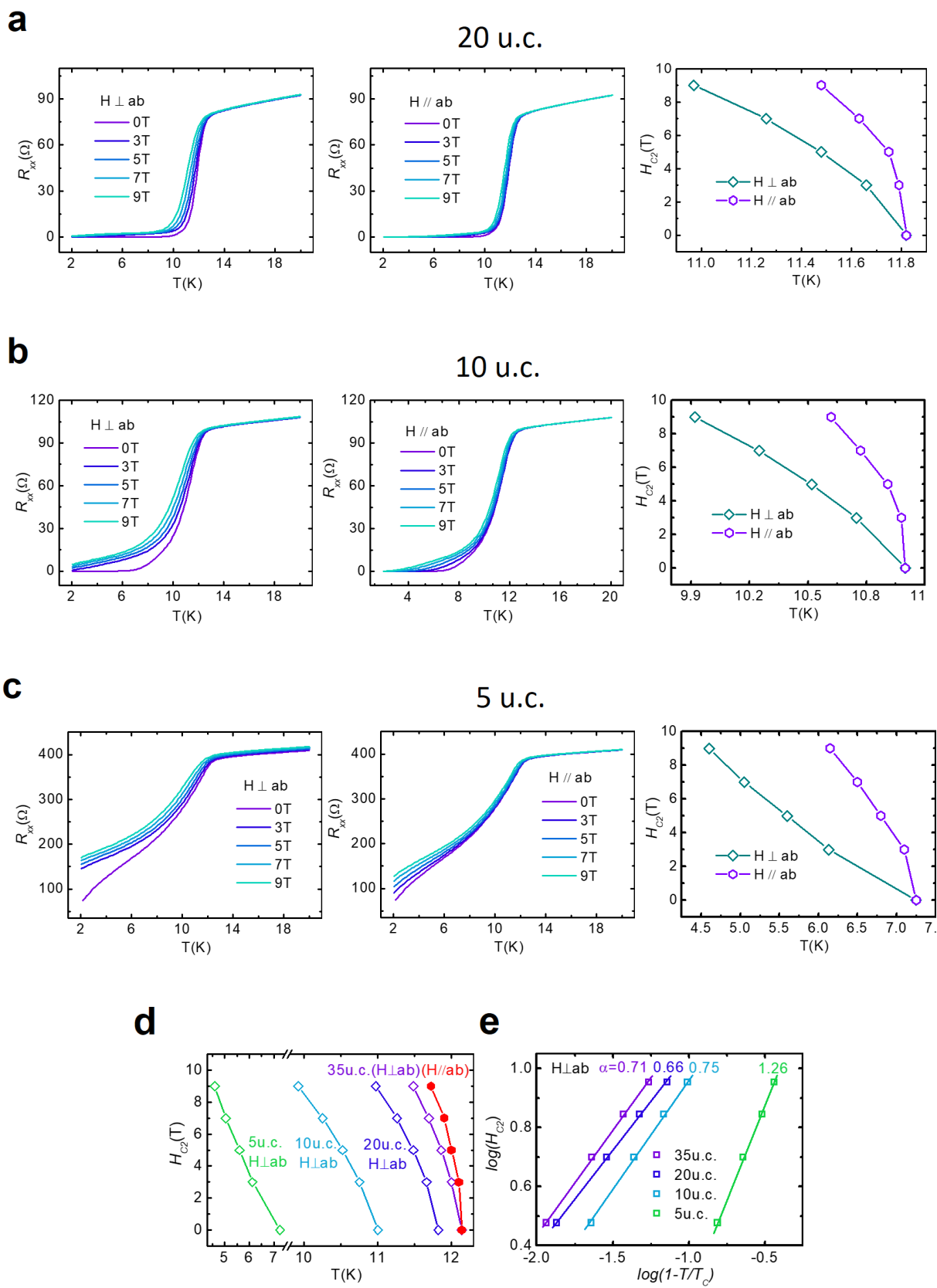

Figure S9: Temperature-dependent longitudinal resistance under varying magnetic fields perpendicular and parallel to the ab plane, and temperature-dependent upper critical fields determined by 50% $R_n$ for (a) 20 u.c., (b) 10 u.c. and (c) 5 u.c. of FeTe$_{0.97}$Se$_{0.03}$ grown on Bi$_2$Te$_3$. (d) $H_{C2}^{\perp}(T)$ of 5 u.c., 10 u.c., 20 u.c. and 35 u.c. FeTe$_{0.97}$Se$_{0.03}$ grown on Bi$_2$Te$_3$ and $H_{C2}^{//}(T)$ of 35 u.c. FeTe$_{0.97}$Se$_{0.03}$/Bi$_2$Te$_3$, determined by 50% $R_n$. (e) $H_{c2}$ fitted by $\left(1 - \frac{T}{T_C}\right)^{\alpha}$, with corresponding $\alpha$ value for each thickness.